%% file: skeleton.tex
\documentclass[a4paper,11pt]{article}
\usepackage{pos}
\usepackage{macros}
\usepackage{dsfont}
\usepackage{subfigure}
\usepackage{booktabs}
\title{Towards the determination of sigma terms for the baryon octet on $N_\mathrm{f} = 2+1$ \textit{CLS} ensembles}
\ShortTitle{Towards the determination of sigma terms for the baryon octet}

\author*[a]{Pia Leonie Jones Petrak}
\author[b]{Gunnar Bali}
\author[b]{Sara Collins}
\author[a]{Jochen Heitger}
\author[b]{Daniel Jenkins}
\author[b]{Simon Weish\"aupl}
\author[b]{Thomas Wurm}

\affiliation[a]{%Institut f\"ur Theoretische Physik,University of M\"unster \\
  Institut für Theoretische Physik, Westfälische Wilhelms-Universität M\"unster,\\
  Wilhelm-Klemm-Straße 9, 48149  M\"unster, Germany}

\affiliation[b]{Institut für Theoretische Physik, Universität Regensburg, 93040 Regensburg, Germany}
\emailAdd{p\_petr04@uni-muenster.de}
\emailAdd{gunnar.bali@ur.de,sara.collins@ur.de}
\emailAdd{heitger@uni-muenster.de}
\emailAdd{Daniel.Jenkins@ur.de}
\emailAdd{Simon.Weishaeupl@ur.de}
\emailAdd{thomas.wurm@ur.de}

\abstract{A lot of progress has been made in the determination of nucleon sigma terms. 
	In this work we consider the sigma terms of the other octet baryons as well.
	These are determined on CLS gauge field ensembles employing the Lüscher-Weisz gluon action and the Sheikholeslami-Wohlert fermion action with $N_\mathrm{f} = 2 + 1$ . The ensembles have pion masses ranging from ${410}\,\mathrm{MeV}$ down to the physical value and lattice spacings covering a range between 
	${0.09}\,\mathrm{fm}$ and ${0.04}\,\mathrm{fm}$. We present some preliminary results 
	for $a\approx 0.06$ fm along a trajectory where the sum of the sea quark masses is kept constant, focusing on the quark mass dependence. We discuss  multi-state fits to tackle the well-known problem of excited state contamination and detail how we analyse connected and disconnected contributions.}

\FullConference{%
 The 38th International Symposium on Lattice Field Theory, LATTICE2021
  26th-30th July, 2021
  Zoom/Gather@Massachusetts Institute of Technology
}

\begin{document}
\maketitle

\input{intro.tex}
\input{main.tex}
%\input{test.tex}
%\begin{thebibliography}{99}
%\bibitem{...}
%....

%\end{thebibliography}
\acknowledgments
\input{acknow.tex}
\bibliographystyle{JHEP}
\setlength{\bibsep}{0pt plus 0.9ex}
\bibliography{bib}

\end{document}

%% file: intro.tex
\section{Introduction}
Sigma terms are referred to as the quark contributions to the mass of a given baryon. They consist of matrix elements of a scalar current $J$ times a quark mass such that
\begin{align}
\sigma_{qB} = m_q \langle B| J |B \rangle
\label{eq:sigma_term}
\end{align}
where $m_q$ denotes the quark mass of flavour $q$. The pion-baryon sigma terms are defined by $\sigma_{\pi B} = \sigma_{uB} + \sigma_{dB}$. We focus on scalar flavour-singlet quark currents $J = \bar{q} \,\mathds{1} \, q$,\, $q\in \{u,d,s\}$. In the matrix element, $B$ refers to the ground state of a baryon $B$.
The most prominent examples are the nucleon sigma terms ($B=N$) which appear in the expressions for WIMP-nucleon scattering cross-sections and are relevant for comparing model predictions
to the exclusion bounds obtained from direct detection dark matter experiments (such as the XENON1T experiment).

We make use of and adjust methods established for the nucleon (reviewed in \cite{Ottnad:2020qbw}) when analysing the entire baryon octet. Studying the sigma terms of the lambda $\Lambda$, sigma $\Sigma$ and cascade $\Xi$ baryons allows us to investigate flavour symmetry breaking in the octet.
In addition, discrepancies between results for the pion-nucleon sigma term from Lattice QCD and phenomenology are still to be resolved (see \cite{FlavourLatticeAveragingGroup:2019iem}, and e.g., \cite{Alexandrou:2019brg,Borsanyi:2020bpd}). In a recent paper, results more consistent with phenomenology were obtained by explicitly including $N\pi$ and $N\pi\pi$ excited states in the analysis \cite{Gupta:2021ahb}. By considering baryons other than the nucleon, we hope to understand the sigma terms in more detail so as to help solve this puzzle. 

%% file: main.tex
\section{Excited state analysis - Ratio method}
\label{sect:ratio_method}
The ratio method \cite{Green:2018vxw,Ottnad:2020qbw} is a way of extracting the ground-state matrix element needed to construct sigma terms (eq.~(\ref{eq:sigma_term})).
We consider the two- and three-point functions of a baryon (from the octet) at rest in the initial and final state.
The spectral decomposition of the two-point function reads
\begin{align}
C_\mathrm{2pt}(\tf)=\sum_{\vec{x}}\left\langle \mathcal{O}_\mathrm{snk}(\vec{x},\tf) \bar{\mathcal{O}}_\mathrm{src}(\vec{0},0) \right\rangle
= \sum_n |Z_n|^2 e^{-E_n \tf}
\end{align}
where $Z_n\propto\langle\Omega|\mathcal{O}_\mathrm{snk}|n\rangle$ is the overlap of the interpolator $\mathcal{O}_\mathrm{snk}$ onto the state $n$ (and $\Omega$ the vacuum state) and $\tf$ the source-sink separation. Summation over spin and colour indices and projection onto
positive parity are implied. These indices become apparent when writing down the operators explicitly. The interpolators for the four octet baryons are
\begin{align}
	\mathcal{O}_\mathrm{snk}^{\alpha,\mathrm{N}} &= \epsilon^{abc} u_{a}^\alpha \left( u_{b}^\beta (C\gamma_5)^{\beta\gamma}d_c^\gamma\right) \quad \text{and} \quad
	\mathcal{O}_\mathrm{snk}^{\alpha,\Lambda} = \epsilon^{abc} s_{a}^\alpha \left( u_{b}^\beta (C\gamma_5)^{\beta\gamma}d_c^\gamma\right),\\
	\mathcal{O}_\mathrm{snk}^{\alpha,\Sigma} &= \epsilon^{abc} u_{a}^\alpha \left( u_{b}^\beta (C\gamma_5)^{\beta\gamma}s_c^\gamma\right) \quad \,\text{and} \quad
	\mathcal{O}_\mathrm{snk}^{\alpha,\Xi} = \epsilon^{abc} s_{a}^\alpha \left( s_{b}^\beta (C\gamma_5)^{\beta\gamma}u_c^\gamma\right).
\end{align}
$a,b,c$ are colour indices, $\alpha,\beta,\gamma$ are spin indices and  $\mathcal{O}_\mathrm{src}^\alpha = \mathcal{O}_\mathrm{snk}^\alpha$ and $\bar{\mathcal{O}}_\mathrm{src} = \mathcal{O}_\mathrm{src}^\dagger \gamma_4 $. $C$ stands for the charge conjugation operator.

Turning to the three-point function, its spectral decomposition reads
\begin{align}
C_\mathrm{3pt}(\tf,t) &=\sum_{\vec{x},\vec{y}}\left\langle \mathcal{O}_\mathrm{snk}(\vec{x},\tf) J(\vec{y},t) \bar{\mathcal{O}}_\mathrm{src}(\vec{0},0) \right\rangle
- \sum_{\vec{x},\vec{y}} \left\langle J(\vec{y},t)\right\rangle\left\langle \mathcal{O}_\mathrm{snk}(\vec{x},\tf)\nonumber \bar{\mathcal{O}}_\mathrm{src}(\vec{0},0) \right\rangle\\
&=\sum_{n,n'} Z_{n'} Z_n^* \langle n'|J|n\rangle
e^{-E_nt} e^{-E_{n'}(\tf-t)},
\end{align}
where $t$ is the insertion time of the scalar current, $J = \bar{q} \, \mathds{1} \, q$,\, $q\in \{u,d,s\}$. % Note that depending on the type of baryon different currents contribute. 
As $J$ has the same quantum numbers as the vacuum, the vacuum expectation value needs to be subtracted. %Also note that both disconnected and connected diagrams need to be evaluated.
Note that depending on the type of baryon, different Wick contractions (so different currents) contribute that result in connected and disconnected quark-line diagrams. 

Taking the ratio of the two spectral decompositions leads to 

\begin{align}
R_\Gamma(\tf,t) = \frac{C_\mathrm{3pt}(\tf,t)}{C_\mathrm{2pt}(\tf)} = g_S^q + c_{01} \mathrm{e}^{-\Delta \, \cdot \,t} + c_{10} \mathrm{e}^{-\Delta \, \cdot \, (\tf-t)} + c_{11} \mathrm{e}^{-\Delta \, \cdot \, \tf} + ...
\label{eq:multi_state_fit_formula}
\end{align}
where $g_S^q = \langle B|J| B\rangle =\langle B|\bar{q} \, \mathds{1} \, q| B\rangle $ is the ground-state matrix element of interest. $\Delta = E_1  - E_0$ is the energy gap between the ground state and the first excited state. The coefficients $c_{01},c_{10},c_{11}$ are made up of matrix elements of different transitions such as  $N_1 \rightarrow N$ , $N \rightarrow N_1$ and
$N_1 \rightarrow N_1$ for the nucleon and similarly for the other three baryons. $N_1$ stands for the first excited state of the nucleon and may be a single- or multi-particle state. As we consider the baryon at rest, $c_{01} = c_{10} \equiv c_1$  holds in this case.
\section{Renormalisation}
Quark masses $m_q$ are renormalised via  
\begin{align}   
m_q^\mathrm{ren} = \zm   \left[m_q \, + \, (\rmsea - 1) \frac{\Tr M}{\Nf} \right],
\end{align}
which holds up to cut-off effects. $\zm$ is the renormalisation parameter of the non-singlet scalar density and $\Tr M = \Sigma_q m_q$.
The matrix elements must renormalise in the inverse manner w.r.t. the masses so that
\begin{align}
\sigma_{qB}^{\mathrm{ren}} = \left(m_q + (\rmsea-1)\frac{\Tr M}{\Nf} \right) \left(g_{q,S}^B + (\rmsea^{-1}-1)\frac{\Tr g_{S}^B}{\Nf}  \right)
\label{eq:renormalisation}
\end{align}
and $\sigma_{\pi B}^\mathrm{ren} = \sigma_{uB}^\mathrm{ren} + \sigma_{dB}^\mathrm{ren}$. The normalisation factor $\rmsea$ is the ratio of flavour non-singlet and singlet scalar density renormalisation parameters, determined in Refs. \cite{Bali:2016umi,Heitger:2021bmg} for our lattice discretisation. It accounts for the mixing of quark flavours under renormalisation for Wilson fermions.
\section{Numerical setup}

\begin{figure}[h]
	\centering
	\includegraphics[width=0.5\linewidth]{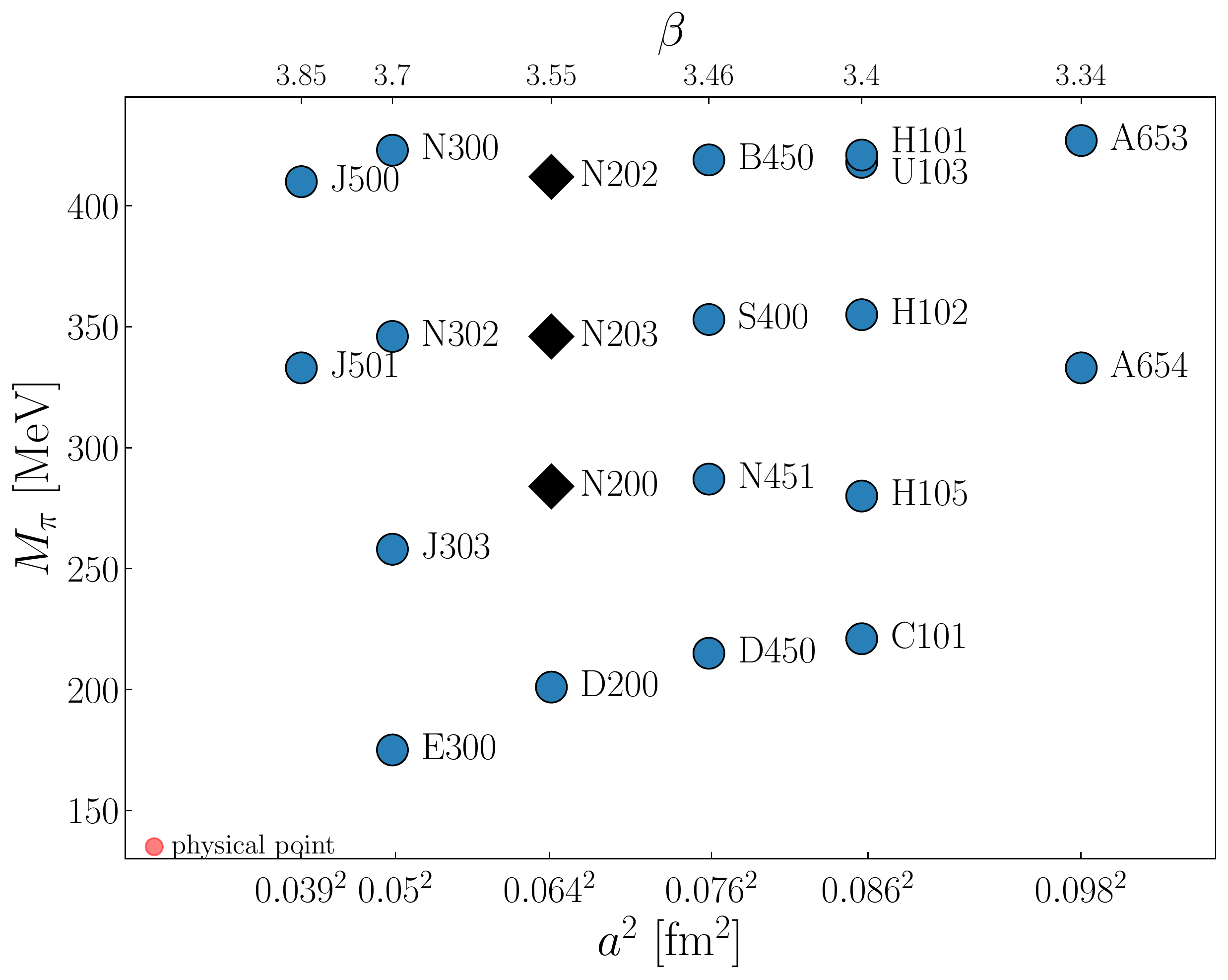}
	\caption{Overview of the $\Tr \mathrm{M}=\mathrm{const.}$ ensembles part of the \textit{CLS} effort that we plan to include in our analysis. So far the ensembles highlighted in black (diamonds) have been analysed. The pion masses and lattice spacings are given on the $y$ and $x $ axes, respectively. The red circle indicates the physical point.}
	\label{fig:ensembles}
\end{figure}

We perform our calculations on \textit{CLS} gauge field ensembles \cite{Bruno:2014jqa} employing the L\"uscher-Weisz gluon action and the Sheikholeslami-Wohlert fermion action with $N_\mathrm{f} = 2 + 1$ ($m_l=m_u=m_d\leq m_s$). Pion-baryon and strange sigma terms are determined on the three ensembles highlighted in (black) diamonds in fig.~\ref{fig:ensembles} along a trajectory where the sum of the sea quark masses is kept constant. Only one lattice spacing of ${0.06426(74)(17)}\,\mathrm{fm}$ ($\beta=3.55$)~\cite{Bruno:2016plf} and a lattice size of $128 \times 48^3$ have been considered so far, focusing on the quark mass dependence. We take three pion masses into account: ${411}\,\mathrm{MeV}$,  ${345}\,\mathrm{MeV}$ and ${284}\,\mathrm{MeV}$. We use $r_\mathrm{m}(\beta=3.55)=1.523(14)$ non-perturbatively determined in \cite{Heitger:2021bmg}.
To compute the connected three-point correlation functions on the $m_l = m_s$ ensemble (N202), we used the standard sequential source method \cite{Maiani:1987by}. On the other ensembles we employed the stochastic method described  in \cite{Bali:2019svt,Bali:2017mft}  (see also \cite{Yang:2015zja,Alexandrou:2013xon,Bali:2013gxx,Evans:2010tg}), estimating a timeslice-to-all propagator. This approach enables us to obtain measurements for all baryons of interest, as multiple source and insertion positions can be estimated simultaneously.
Four different source-sink separations, $\tf/a =  [11, 14, 16, 19]$, corresponding to $\tf \approx [0.71 \,\mathrm{fm}, 0.9 \,\mathrm{fm}, 1.03 \, \mathrm{fm}, 1.22 \, \mathrm{fm}]$, are employed. Four measurements ($2$ replica $\times$ (forward and backward direction)) are performed for each $\tf$ on every configuration except for the $m_l = m_s$ ensemble (N202) where we used the sequential source method; here, only one measurement is undertaken at $\tf=11$ and two at $\tf=[14,16]$ (whereas the number of measurements is also four at $\tf=19$).

The disconnected three-point functions are constructed by correlating a quark loop with a baryon two-point function. The loop is estimated stochastically leading to additional noise on top of the Monte-Carlo gauge sampling. In order to reduce the noise, the truncated solver method~\cite{Bali:2009hu}, the hopping parameter expansion technique~\cite{Thron:1997iy} and time partitioning~\cite{Bernardson:1993he} are utilised. Forty measurements ($2$ replica $\times$ $20$ different spatial source positions) of the two-point function are performed on each configuration with the exception of N202 where the number is 52. % this leads to the same number of measurements for the disconnected three-point functions.
The source-sink separations range from $\tf/a=4  \leftrightarrow \tf \approx 0.26\,\mathrm{fm}$ to $\tf/a=19 \leftrightarrow \tf \approx 1.22\,\mathrm{fm}$.
For the analysis of the statistical errors we employ the $\Gamma$-method \cite{Wolff:2003sm} that is based on autocorrelation functions. %We plan to but do not yet account for the remaining critical slowing down of the Monte Carlo algorithm by attaching a tail to the autocorrelation function, as suggested in ref.~\cite{Schaefer:2010hu}. 

%\begin{itemize}
	%\item preliminary (without errors) values used: pion/kaon masses
	%\item preliminary values with error used: $\kappa_\mathrm{crit}$, $\rmsea$ - part of table?
	%lattice spacing a
	%\item t_0(M_pi,M_k)
%\end{itemize}
\section{Analysis and preliminary results}

%\begin{itemize}
	%\item show ratios for baryons on N203 that differ
	%\item Leading order expectation from SU(3) flavour symmetry - ChPT formulas? NO
	%\cite{Tiburzi:2004rh,Bernard:2005fy,MartinCamalich:2010fp}?´
%\end{itemize}

%The ratio, see sect.~\ref{sect:ratio_method} is determined on each of the three ensembles. The connected and disconnected contributions are kept separate. The connected and disconnected ratios each involve separate ratios for different currents. In order to tackle excited state contamination we perform multi-state fits, according to eq.~\ref{eq:multi_state_fit_formula}. For each baryon we fit all connected and disconnected ratios simultaneously the energy gap $\Delta$ being the common fit parameter. This way it is possible to constrain the energy gap which is one of the challenges. From the simultaneous fit we obtain the ground-state matrix elements of interest. They are combined and multiplied by the according quark masses as to make up strange and pion sigma terms for all of the baryon octet. {\color{red} less detail?}

\begin{figure}[h]
	\includegraphics[width=0.5\linewidth]{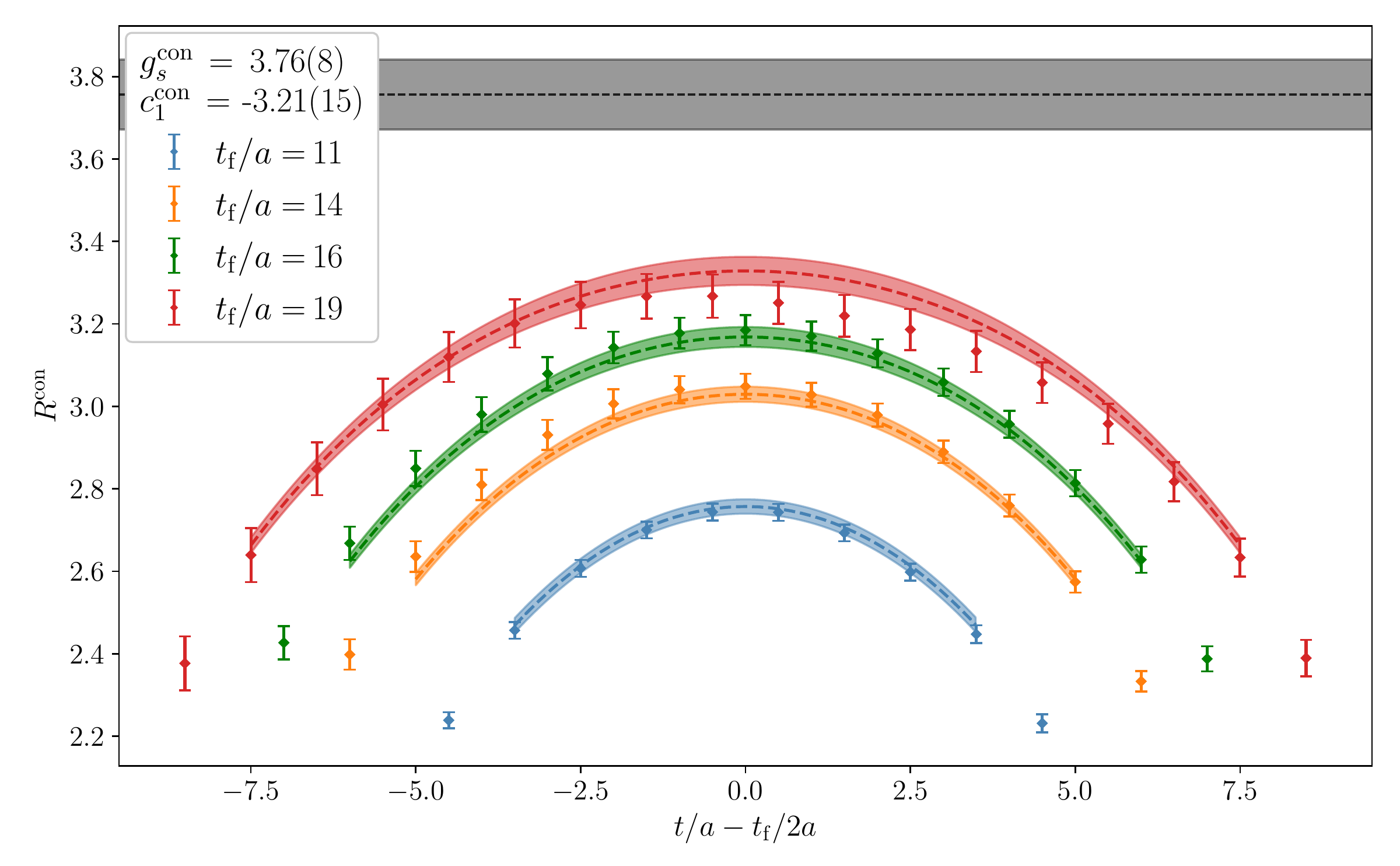}
	\includegraphics[width=0.5\linewidth]{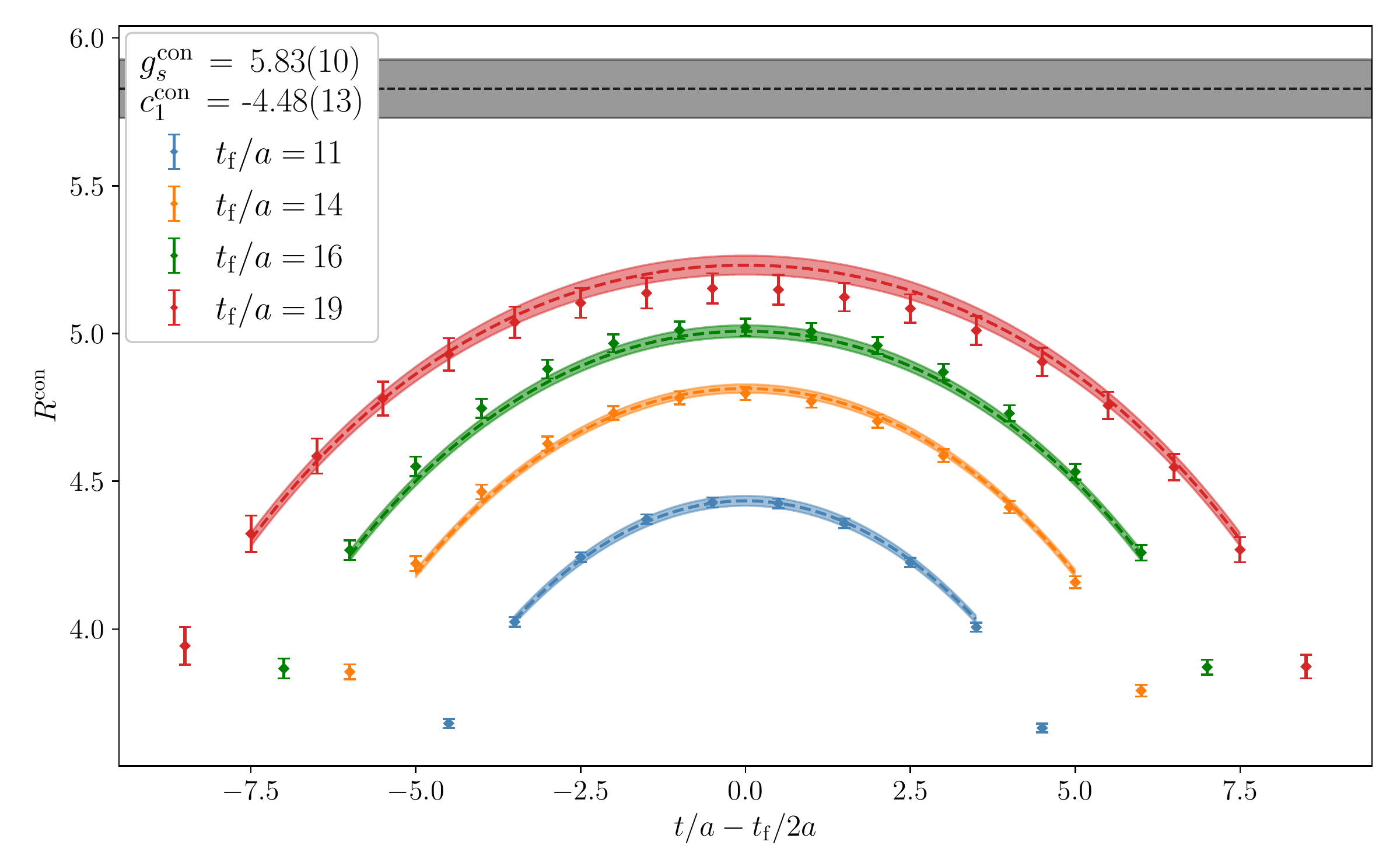}
	\includegraphics[width=0.5\linewidth]{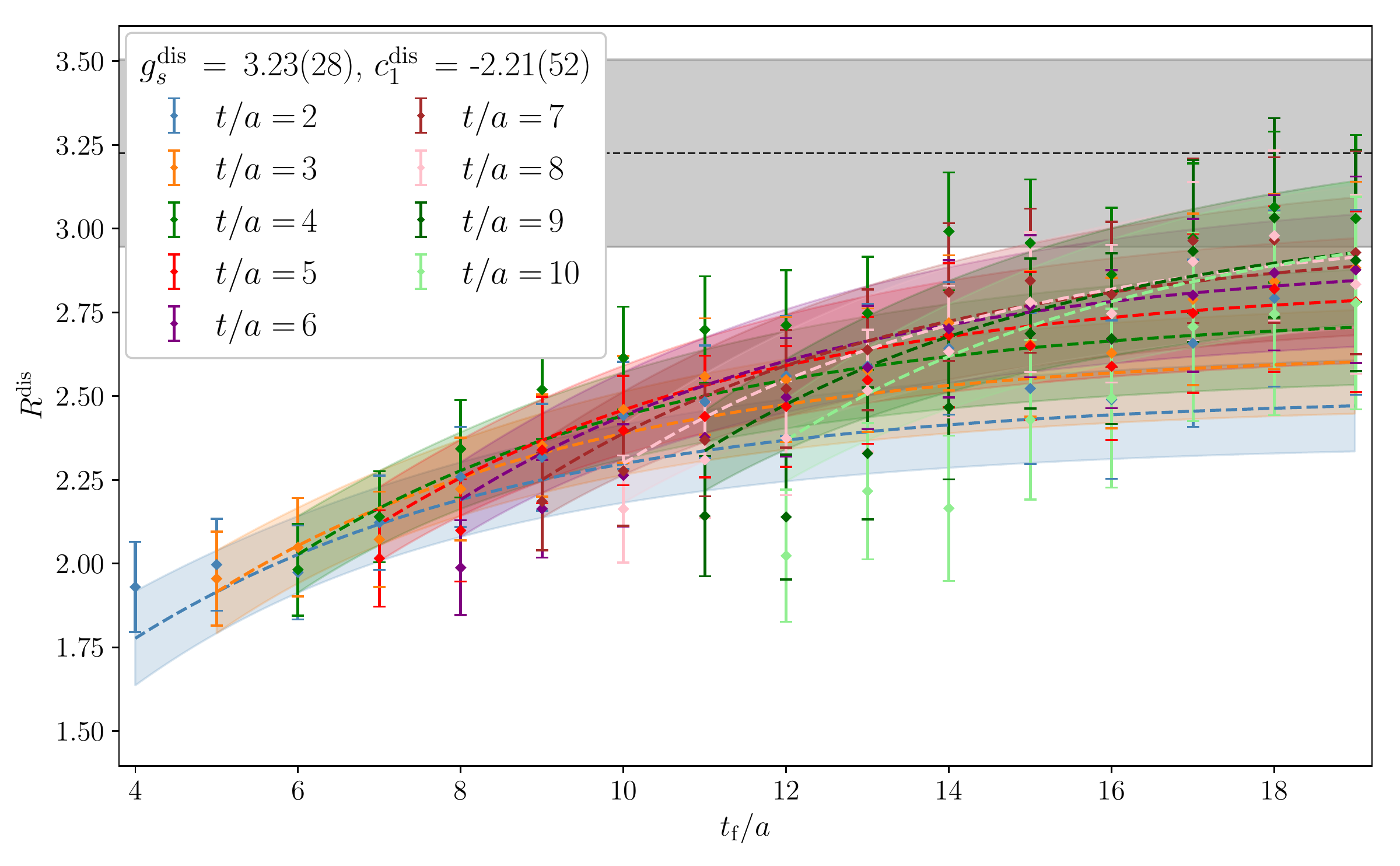}
	\includegraphics[width=0.5\linewidth]{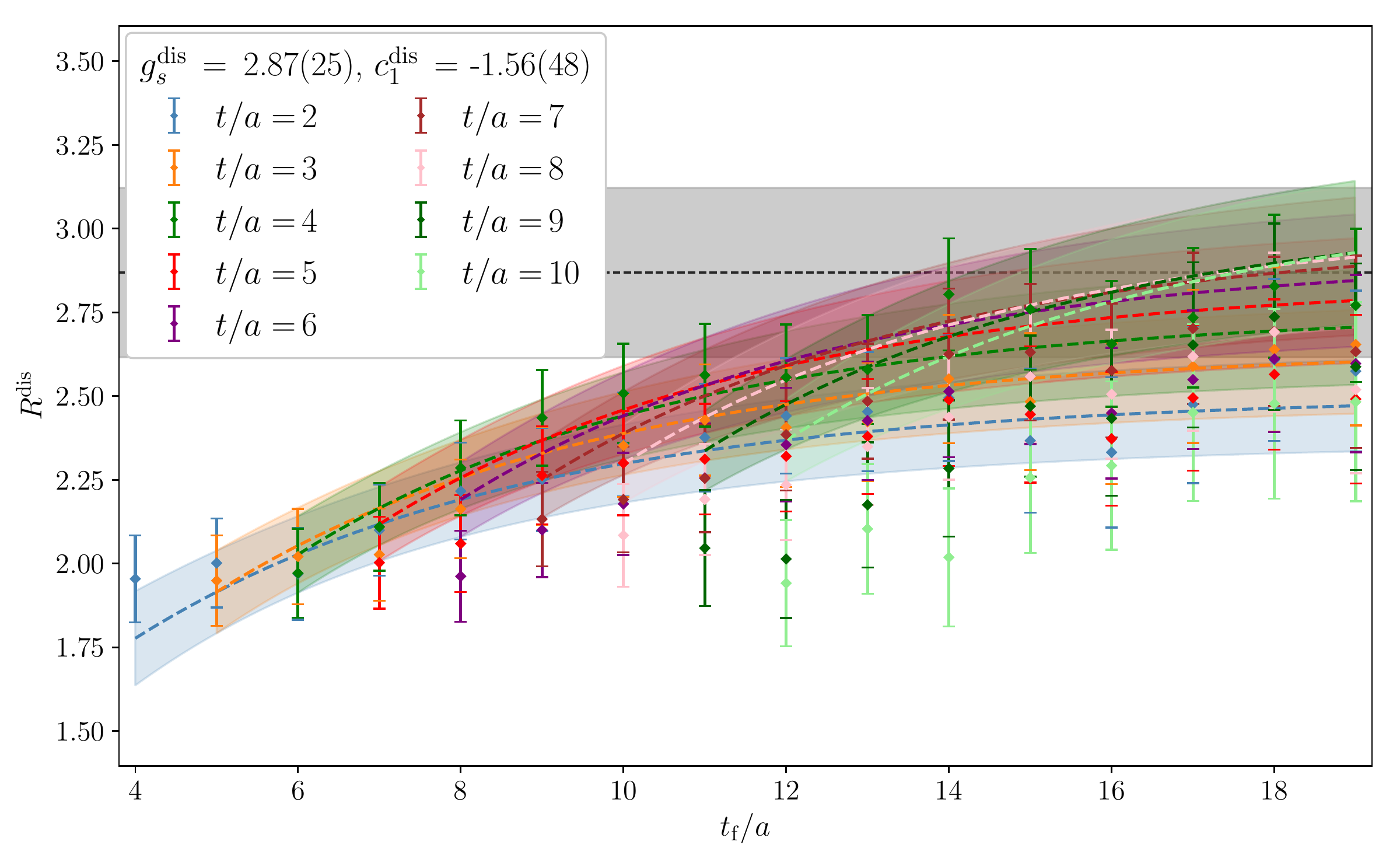}
	\caption{The connected and disconnected ratios that contribute to the sigma terms of the $\Xi$ baryon for ensemble N203: Simultaneous fit to the connected and disconnected ratios is indicated by the coloured shaded regions, with the resulting ground state
	scalar matrix element displayed as a grey band.  We obtain $\chi^2/\mathrm{d.o.f.} = 0.72$ and an energy gap of $\Delta \approx 651\, \mathrm {MeV}$. %0.2121(95). 
	At the top, the connected ratios are plotted against the insertion time $t$ at different source-sink separations $\tf$  for the $\bar{u}u$ current (left) and the $\bar{s}{s}$ current (right). At the bottom, the disconnected ratios are plotted against the source-sink separation $\tf$ at different insertion times $t$ for $J=\bar{l}l$ (left) and $J=\bar{s}s$ (right) .}
	\label{fig:sim_fit_lambda}
\end{figure}

Connected and disconnected ratios are constructed separately for all scalar currents that contribute. In order to tackle excited state contamination we perform multi-state fits, according to eq.~(\ref{eq:multi_state_fit_formula}). For each baryon we fit all connected and disconnected ratios simultaneously, with the energy gap $\Delta$ being the common fit parameter.  %{\color{green}the energy gap should be independent of the current, so this is one way to constrain it.}
As an example, the ratios (and fits) relevant for determining the sigma terms of the $\Xi$ baryon on the N203 ensemble are displayed in fig.~\ref{fig:sim_fit_lambda} showing all ratios involved. While we were able to resolve the first two excited state terms from eq.~(\ref{eq:multi_state_fit_formula}), it was not possible to resolve the third and we set $c_{11}=0$ throughout our analysis. The $\chi^2/\text{d.o.f}$ values were below one for all baryons.
\begin{figure}[h]
	\centering
	\includegraphics[width=0.8\linewidth]{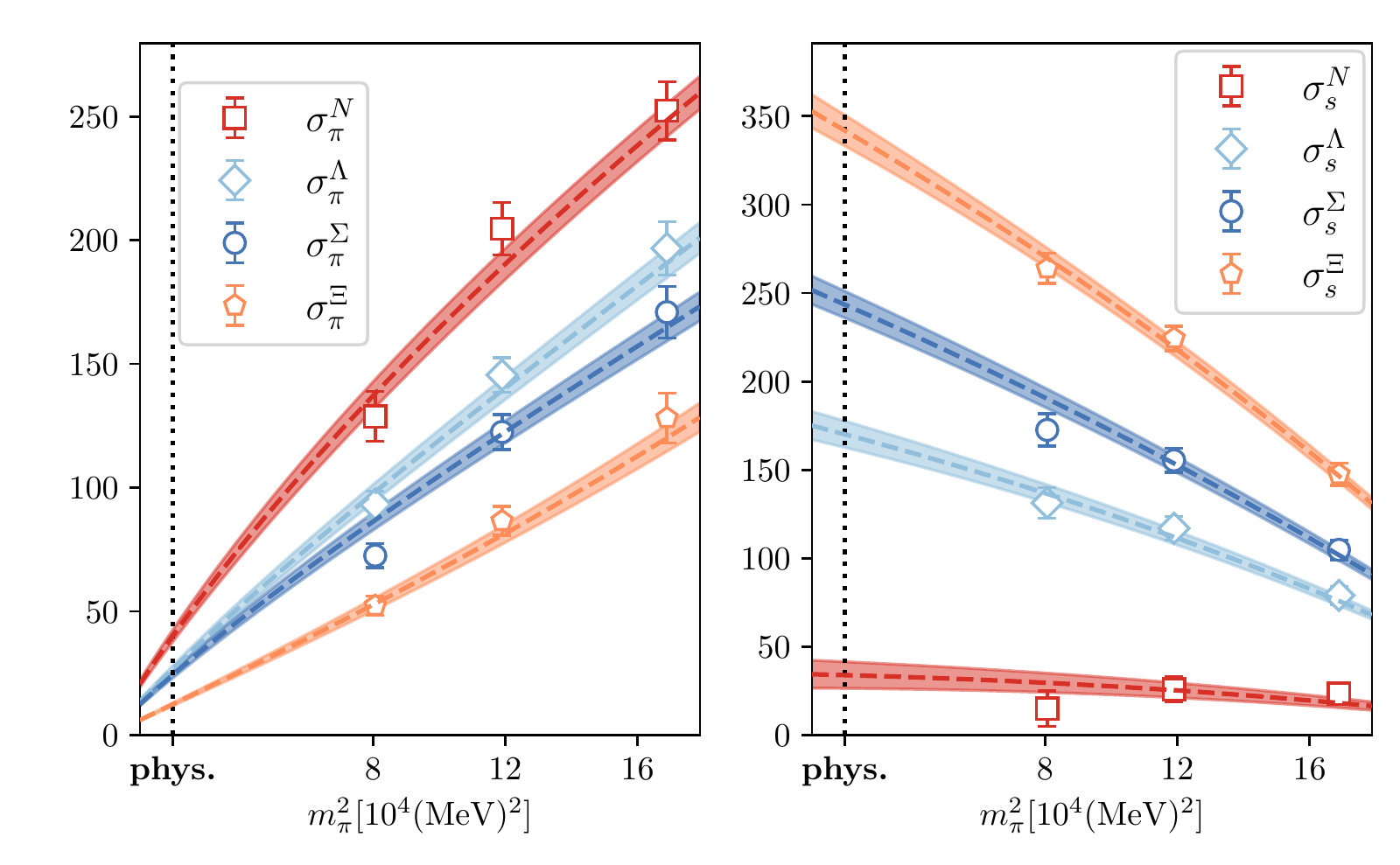}
	\caption{Pion mass dependence of sigma terms: The dotted vertical lines point to the physical pion mass. The pion-baryon (left) and strange-baryon (right) terms are depicted by squares (nucleon), diamonds ($\Lambda$), circles ($\Sigma$) and pentagons ($\Xi$). Simultaneous fit to pion-baryon and strange sigma terms is displayed by the dashed lines (including the error bands as shaded regions) resulting in $\chi^2/\mathrm{d.o.f}=1.29$. Both LO LECs and the octet baryon mass in the chiral limit are kept fixed to $F = 0.446(7)$, $D = 0.731(12)$ and  $m_0 = 729(42)\, \mathrm{MeV}$ from a preliminary analysis of the nucleon mass and the axial charges in the chiral limit; while the NLO LECs $b_D,b_F,\bar{b}$ and the pion decay constant $F_0$ are fitted. We get $\bar{b}=0.00317(29)$, $b_F=-0.000335(27)$, $b_D=0.0000493(21)$ and $F_0=119.9(9.8)\, \mathrm{MeV}$. It differs greatly from  $F_0 = 71(2)\, \mathrm{MeV}$, the preliminary value from a combined fit to the pion decay constant and the pion mass part of the analysis mentioned above.}
	\label{fig:chpt}
\end{figure}
The ground-state matrix elements of interest can now be extracted from the fit. Matrix elements of different currents are combined and multiplied by the corresponding quark masses as to make up pion-baryon and strange sigma terms for all octet baryons considered, see eq.~(\ref{eq:sigma_term}). Renormalisation is applied  via eq.~(\ref{eq:renormalisation}). Our preliminary results for pion-baryon and strange sigma terms are plotted against the pion mass in fig.~\ref{fig:chpt}. From Baryon Chiral Perturbation Theory (BChPT) we can derive the pion mass dependence expected from SU(3) flavour symmetry \cite  {Lehnhart:2004vi} (see also \cite{PhysRev.125.1067,10.1143/PTP.27.949,Geng:2013xn}); we
apply the Feynman-Hellmann theorem that relates sigma terms to derivatives of the baryon mass with respect to quark masses, resulting in

\begin{subequations}
	\label{eq:chpts}
	\begin{equation}
	\sigma_{\pi{}B}=M_{\pi}^2\left\{\frac{2}{3}\bar{b}-\delta b_B
	+\frac{m_0^2}{(4\pi F_0)^2}
	\left[\frac{g_{B,\pi}}{2M_{\pi}}f'\left(\frac{M_{\pi}}{m_0}\right)
	+\frac{g_{B,K}}{4M_{K}}f'\left(\frac{M_{K}}{m_0}\right)
	+\frac{g_{B,\eta}}{6M_{\eta}}f'\left(\frac{M_{\eta}}{m_0}\right)
	\right]\right\},\label{eq:cpt1}
	\end{equation}
	\begin{equation}
	\sigma_s=\left(2M_K^2-M^2_{\pi}\right)
	\left\{\frac{1}{3}\bar{b}+\delta b_B
	+\frac{m_0^2}{(4\pi F_0)^2}
	\left[
	\frac{g_{B,K}}{4M_{K}}f'\left(\frac{M_{K}}{m_0}\right)
	+\frac{g_{B,\eta}}{3M_{\eta}}f'\left(\frac{M_{\eta}}{m_0}\right)
	\right]
	\right\},\, \quad \quad  \label{eq:cpt2}
	\end{equation}
\end{subequations} 
where $m_0$ and $F_0$ are the octet baryon mass and pion decay constant in the chiral limit. $\delta b_B$ is a combination of two of the three BChPT next-to-leading order (NLO) low energy constants (LECs) $b_D,b_F,\bar{b}=-6b_0-4b_D$ and depends on the baryon,
\begin{align}
\delta b_N=\tfrac23(3b_F-b_D),\quad
\delta b_\Lambda=-\tfrac43 b_D,\quad 
\delta b_\Sigma=\tfrac43 b_D,\quad 
\delta b_\Xi=-\tfrac23(3b_F+b_D).\label{eq:deltab}
\end{align}

\noindent The couplings $g_{B,\pi},g_{B,K}$ and $g_{B,\eta_8}$ are made up of different combinations of the leading order (LO) LECs $F$ and $D$,
\begin{align}
g_{N,\pi}&= \tfrac{3}{2}{(D+F)}^2, & g_{N,K}&=\tfrac{5}{3} D^2 - 2D F + 3 F^2, & g_{N,\eta_8} &= \tfrac{1}{6} {(D-3F)}^2,\nonumber \\
g_{\Lambda,\pi}&=2 D^2, & g_{\Lambda,K}&=\tfrac{2}{3} D^2 + 6 F^2, & g_{\Lambda,\eta_8}&=\tfrac{2}{3} D^2,\nonumber \\
g_{\Sigma,\pi}&=\tfrac{2}{3}D^2 + 4 F^2, & g_{\Sigma,K}&=2D^2 + 2F^2, & g_{\Sigma,\eta_8}&=\tfrac{2}{3} D^2,\nonumber \\
g_{\Xi,\pi}&=\tfrac{3}{2}{(D-F)}^2, & g_{\Xi,K}&=\tfrac{5}{3} D^2 + 2D F + 3 F^2, & g_{\Xi,\eta_8}&=\tfrac{1}{6} {(D+3F)}^2,
\end{align}
that also appear in the ChPT expressions for the axial charges.
$f^\prime$ is the derivative of the loop function $f$ that is set to $f(x)=-\pi x^3$ in Heavy Baryon ChPT \cite{Bernard:1992qa,Gasser:1987rb} or 
\begin{align}
f(x)= -2x^3\left[\sqrt{1-\frac{x^2}{4}}\arccos\left(\frac{x}{2}\right)
+\frac{x}{2}\ln(x)\right].
\label{eq:loop_fct_BChPT}
\end{align}
in covariant BChPT in the extended on-mass-shell (EOMS) scheme  \cite{Gegelia:1999gf,Fuchs:2003qc,Lehnhart:2004vi}.
This BChPT prediction (\ref{eq:chpts}) tells us that pion-baryon and strange sigma terms should be describable by the same set of LECs. We find that fitting our preliminary pion-baryon and strange sigma terms simultaneously is successful so we can describe both sigma terms consistently, see fig.~\ref{fig:chpt}.

In addition, we investigate whether we obtain consistent results for the LECs with a preliminary study where $F_0$ was estimated from a combined fit to the pion decay constant and the pion mass. As part of the same study $m_0$, $F$ and $D$ were determined in an analysis of the nucleon mass and the axial charges in the chiral limit.
%In addition, we investigate whether the NNLO LECs from a preliminary PCAC mass analysis can be obtained. 
We see that it is not possible to arrive at a satisfactory fit keeping these four parameters fixed to the preliminary values. Instead we find that at least one parameter has to account for the difference in curvature. %and  finding a  consistent set of ChPT LECs that describes meson decay constants, axial charges
%and sigma terms is a remaining challenge.
We show the best fit for our sigma terms in fig.~\ref{fig:chpt}; we perform a simultaneous fit to pion-baryon and strange sigma terms according to eq.~(\ref{eq:chpts}) using eq.~(\ref{eq:loop_fct_BChPT}) for the loop function. Here the three NLO LECs and $F_0$ are the common fit parameters whilst keeping $F$, $D$ and $m_0$ fixed to the values from the aforementioned (preliminary) analysis. Our fit result for $F_0$ is unreasonably large. This may be due to the fact that we do not yet incorporate cut-off and finite-volume effects on this small subset of ensembles at a single lattice spacing. Note that higher order ChPT effects may also contribute.

%\begin{figure}
%	\includegraphics[width=1.0\linewidth]{figures/chiral_extrapolation.pdf}
%	\caption{Pion mass dependence of the sigma terms. The next-to leading order expectation from SU(3) flavour symmetry  \cite{PhysRev.125.1067,10.1143/PTP.27.949,Geng:2013xn} is depicted; we only show lines to guide the eye fixed by the symmetric point (N202, the rightmost points).}
%	\label{fig:chpt}
%\end{figure}

%\begin{figure}
%	\includegraphics[width=1.0\linewidth]{figures/sigma_ChPT_fit_pion_linear_strange_NNLO.pdf}
%	\caption{\color{red} Pion mass dependence of the sigma terms. The expectation from SU(3) flavour symmetry  \cite{PhysRev.125.1067,10.1143/PTP.27.949,Geng:2013xn} is depicted; for the pion sigma terms (left) the uncorrelated simultaneous fit to all baryons worked best when excluding the NNLO contribution ($\chi^2/\mathrm{d.o.f}=1.1$) whereas we were able to also resolve NNLO contributions ($\chi^2/\mathrm{d.o.f}=1.68$)  for the strange sigma terms (right) using the loop function from eq.~(\ref{eq:loop_fct_BChPT}). Note that we used the preliminary values, $F_0 = 71 \mathrm{MeV}$, $m_0 = 729 \mathrm{MeV}$, $F = 0.446$ and $D = 0.731$ presented by Simon Weishäupl at \emph{https://indi.to/4D22k} ( Would I actually quote this or not give the actual values?).\\
%	{\color{red} $b_D,b_F,\bar{b}$ for:\\
%	Pion: [0.001629(58)], [-0.0002180(80)], [0.0000783(44)]\\
%	Strange: [0.006274(48)], Obs[-0.0006254(36)], Obs[0.0000264(25)]\\
%	'compatible' with values fixed by symmetric point at NLO (so only for pion fit):\\
%	$[0.001632, -0.000187, 0.0000582]$}}
%\end{figure}

\section{Conclusion and outlook}
We have demonstrated that it is possible to obtain  pion-baryon and strange sigma terms for all octet baryons using similar methods to those for the nucleon. Taking a closer look at the renormalisation pattern, it might be more convenient to consider other combinations of sigma terms. We also aim to take into account all main sources of systematics. We will for example try out further fitting techniques; in order to determine whether we control excited state contributions sufficiently, the summation method \cite{Green:2018vxw,Ottnad:2020qbw} may serve as a cross-check. In the future we plan to extend the analysis to include additional ensembles. This will allow for a chiral extrapolation to the physical pion mass and an investigation of cut-off and finite-volume effects.

%% file: acknow.tex
This work is supported by the Deutsche Forschungsgemeinschaft (DFG) through the Research Training Group ``GRK 2149: Strong and Weak Interactions -- from Hadrons to Dark Matter'' (P. L. J. P. and J. H.). %We acknowledge the computer resources provided by the WWU IT, formerly ‘Zentrum für Informationsverarbeitung (ZIV)’, of the University of Münster (PALMA-II HPC cluster) and thank its staff for support.
G. B., S. C., D. J., S. W. and T. W. were supported by the European Union’s Horizon 2020 research and innovation programme under the Marie Skłodowska-Curie grant agreement no. 813942 (ITN EuroPLEx) and grant agreement no. 824093 (STRONG-2020).\\ 
We gratefully acknowledge computing time granted by the
John von Neumann Institute for Computing (NIC), provided on the Booster
partition of the supercomputer JURECA~\cite{jureca} at
\href{http://www.fz-juelich.de/ias/jsc/}{J\"ulich Supercomputing Centre (JSC)}.
Additional simulations were carried out at the QPACE~3
Xeon Phi cluster of SFB/TRR~55.
The authors also gratefully acknowledge the Helmholtz Data Federation (HDF) for funding this work by providing services and computing time on the HDF Cloud cluster at the Jülich Supercomputing Centre (JSC)~\cite{hdfcloud}.